\documentclass[traditabstract]{aa}

\usepackage{txfonts}
\usepackage{natbib}
\usepackage{graphicx}
\usepackage{url}

\def\aj{AJ}%
\def\apj{ApJ}%
\def\apjl{ApJ}%
\def\apjs{ApJS}%
\def\aap{A\&A}%
\def\mnras{MNRAS}%
\def\pasp{PASP}%
\def\pasj{PASJ}%
\def\nat{Nature}%
\def\s4g{S$^4$G}

\def\hi{H{\sc i}}
\def\mum{$\mu$m}

\begin{document}

\title{Herschel/SPIRE  Observations of the Dusty Disk of NGC 4244 \thanks{{\it Herschel} is an ESA space observatory with science instruments provided by European-led Principal Investigator consortia and with important participation from NASA.}}
 
\subtitle{}
\authorrunning{Holwerda et al.}
\titlerunning{Herschel Observations of NGC 4244}

\author{B.W. Holwerda\inst{1}, S. Bianchi\inst{2}, T. B\"{o}ker\inst{1}, D. Radburn-Smith\inst{3},  R. S. de Jong\inst{4}, M. Baes\inst{5}, P.C. van der Kruit\inst{6}, M.Xilouris\inst{7}, K.D. Gordon\inst{8}, and J.J. Dalcanton\inst{3}   }

\offprints{B.W. Holwerda, \email{benne.holwerda@esa.int}}

\institute{
$^1$ European Space Agency Research Fellow (ESTEC), Keplerlaan 1, 2200 AG Noordwijk, The Netherlands \\
$^2$ INAF-Arcetri Astrophysical Observatory, Largo Enrico Fermi 5, I - 50125 Florence, Italy\\
$^3$ Department of Astronomy, University of Washington, Box 351580, Seattle, WA 98195, USA\\
$^4$ Leibniz Institut f\"ur Astrophysik Potsdam (AIP), An der Sternwarte 16, 14482 Potsdam, Germany\\
$^5$ Sterrenkundig Observatorium, Universiteit Gent, Krijgslaan 281-S9, B-9000 Gent, Belgium \\
$^6$ Kapteyn Astronomical Institute, University of Groningen, P.O. Box 800, 9700 AV Groningen, the Netherlands\\
$^7$ Institute of Astronomy and Astrophysics, National Observatory of Athens, P. Penteli, 15236 Athens, Greece\\
$^8$ Space Telescope Science Institute, 3700 San Martin Drive, MD 21218, Baltimore, USA}

\date{Received  9-12-2011 / Accepted YYY}

% \abstract {context} {aims} {method} {results} {conclusions}
\abstract{
We present Herschel/SPIRE images at 250, 350, and 500 $\mu$m of NGC 4244, a typical low-mass, disk-only and edge-on spiral galaxy. 
The dust disk is clumpy and shows signs of truncation at the break radius of the stellar disk. This disk coincides with the densest part of the \hi \ disk. 
We compare the Spectral Energy Distribution, including the new SPIRE fluxes, to 3D radiative transfer models; a smooth model disk and a clumpy model with embedded heating. Each model requires a very high value for the dust scale-length ($h_d = 2-5~ h_*$), higher dust masses than previous models of NGC 4244 ($M_{d} = 0.47-1.39 \times 10^7 M_\odot$) and a face-on optical depth of $\tau_V^{f.o.} = 0.4-1.12$, in agreement with previous disk opacity studies. The vertical scales of stars and dust are similar. 
The clumpy model much better mimics the general morphology in the sub-mm images and the general SED.
% implications
The inferred gas-to-dust mass ratio is compatible with those of similar low-mass disks. The relatively large radial scale-length of the dust disk points to radial mixing of the dusty ISM within the stellar disk. The large vertical dust scale and the clumpy dust distribution of our SED model are both consistent with a scenario in which the vertical structure of the ISM is dictated by the balance of turbulence and self-gravity.}


\maketitle

\section{Introduction}
\label{s:intro}

The edge-on perspective of a spiral galaxy's disk reveals both the vertical structure of the disk and its faint outskirts, thanks to line-of-sight integration. 
One typical feature is the mid-plane dust absorption lane. Dust is linked to the cold molecular ISM, mechanically \citep{Allen86, Weingartner01b}, through mutual shielding, and as a catalyst for molecular gas formation. 
\cite{Dalcanton04} used the appearance of dust lanes as a probe of vertical stability of spiral disks. They found that in massive spirals, the ISM collapses into a thin dust-lane, while in low-mass disks ($\rm v_{rot} < 120 ~ km/s$) the dust morphology is flocculant. 

One comprehensive approach to investigate this phenomenon is to model the multi-wavelength information on a range of spiral disks with a Spectral Energy Distribution (SED) model. %s are often tested on edge-on spirals case. 
Dust absorption, especially within the thin dark dust lane, is strongest at optical and ultra-violet wavelengths and the challenge for the models is to balance the observed absorption with the dust emission at far-infrared (FIR) and sub-millimeter wavelengths \citep[e.g.,][]{Popescu00, Misselt01, Gordon01, Alton04, Bianchi08, Baes10a, Bianchi11}. % Xilouris97, Xilouris98, Baes11
Until recently, these models were degenerate in the vertical distributions of stellar light and ISM due to a lack of resolution and wavelength coverage in the FIR and sub-millimeter, but with the advent of the {\em Herschel Space Observatory}, the vertical structure of nearby edge-on disks can now be fully resolved.
% {\bf We discussed this many times, and I agree that it could be done with PACS, but here we have SPIRE only...}

Massive edge-on spiral galaxies have been observed by several Herschel surveys \citep[see][]{Boselli10, Eales10, Davies10a} and a dedicated program\footnote{HERschel Observations of Edge-on Spirals, HEROES} (Verstappen et al. {\em in prep.}) which targets seven massive spirals. Our NHEMESES\footnote{New HErschel Multi-wavelength Extragalactic Survey of Edge-on Spirals \citep[][]{Holwerda11iau}} program  aims to observe a complementary sample of low-mass nearby edge-ons. 
The combined observations will serve as a test for the different radiative transfer models in the literature;
% can be explored by several different radiative transfer models in a future benchmark SED study; 
% SKIRT \citep{Baes03, Baes11}, TRADING \citep{Bianchi07, Bianchi08}, DIRTY \citep{Gordon01, Misselt01, Pierini04} and the model from \cite{Popescu00, Popescu11}. These models will further test the suggested dichotomy in the vertical dust structure.
those from \cite{Baes03, Baes11}, \cite{Bianchi07, Bianchi08}, \cite{Gordon01, Misselt01} and \cite{Pierini04}, and \cite{Popescu00, Popescu11}. These models will further test the suggested dichotomy in the vertical dust structure.

Here, we explore the first NHEMESES results on NGC 4244, the prototypical low-mass ($\rm v_{rot} = 95 ~ km/s$), late-type (Sc), edge-on spiral at a distance\footnote{The mean of the distances in NED from \cite{Heald11a}. \cite{GHOSTS} derive a distance of 4.7 Mpc.} of 4.4 Mpc to compare to the first {\em Herschel} results on the massive edge-on spiral NGC 891 \citep{Bianchi11}.
Originally, \cite{vdKruit81a} found a single edge-on exponential disk truncated at 9 kpc, later confirmed by \cite{Fry99}. The truncation was also observed with the {\em Hubble Space Telescope} \citep{de-Jong07}.
%, and \cite{van-der-Kruit07} notes that the HI warp starts shortly beyond this truncation, linking the stellar and ISM phenomena.
% thick disk
\cite{Yoachim06} note a second, thick stellar disk as doubtful, but \cite{Comeron11a} report both a thin and thick disk of equal mass.
Its center hosts a rotating nuclear star-cluster \citep{Seth08}, and the disk is classified as ``corrugated" \citep{Florido91a}.

% ISM
NGC4244's ISM is mostly in a thin disk, both the ionized gas \citep{Hoopes99} and warm dust, implying a low specific star-formation rate \citep{Kodaira96}. The \hi \ disk shows a clear warp, kinematic evidence for a compact dark matter halo, and lagging extra-planar gas \citep
{Olling96b,Zschaechner11b, Heald11b}. % {Olling96b,Olling96a, 
The $H_2$ mass, based on CO observations, is $1.4\times10^7 ~ M_\odot$ \citep{Matthews01}. % , Sage93

A recent SED model by \cite{MacLachlan11} finds that most (80\%) of the FIR emission could arise from an optically thin disk with a similar scale-height for both dust and stars \citep[as did][based on stellar populations]{Seth05a}, a dust scale-length 1.8 times the stellar one and a dust mass of $ 2.38 \times 10^6 M_\odot$. They note their model diverges from the observed disk in the FIR and that the dust model is still uncertain without sub-mm observations. 

In this Letter, we present new SPIRE sub-mm observations of NGC 4244, adding spatial information on the cold dust, and present two TRADING models to illustrate the role of ISM geometry on the observed overall SED and the sub-mm images.

\begin{figure*}
\sidecaption
\includegraphics[width=0.8\textwidth]{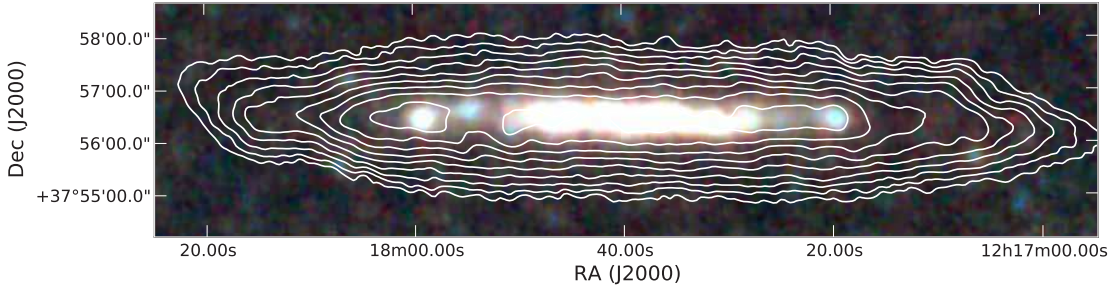}
\caption{A colour composite of the  SPIRE maps of NGC 4244, 250 (red), 350 (green), and 500 $\mu$m (blue) with the \hi\ contours from \cite{Zschaechner11b} (contours at $6.4 \times 10^{19}$ cm$^{-2}$ increasing by factors of 2. The SPIRE flux stays within the $8.1 \times 10^{19}$ cm$^{-2}$ contour (second to highest). }
\label{f:hi}
\end{figure*}

\section{Observations}
\label{s:obs}

SPIRE \cite[Spectral and Photometric Imaging REceiver,][]{Griffin10} instrument at 250, 350 and 500 $\mu$m onboard the ESA {\em Herschel Space Observatory} \citep{Pilbratt10a} observed NGC 4244 at 250, 350 and 500 $\mu$m  in large map mode, over 25'x25' centered on the object with two cross-scans, using a 30" s$^{-1}$ scan rate. 
Data-reduction is as in \cite{Bianchi11}; we generated maps with the na\"{i}ve map-making procedure within HIPE \citep{Ott10}, with pixel sizes of 6", 8", 12" at 250, 350 and 500 \mum, respectively (about 1/3 of the measured FWHM: $\approx$ 18", 25" and 36"), resulting in uniform backgrounds and sky rms noise of 1, 0.6 and 0.3  MJy sr$^{-1}$ respectively, Fig. \ref{f:hi}).
A color correction half-way between that for a point- and extended-source was implemented with a color-correction in between the point- and extended-source (marginally resolved vertically), we derive integrated fluxes of 15.9, 9.8, and 5.1 Jy (250, 350 and 500 \mum), and adopt a conservative 15\% calibration error.

\section{Results \& Analysis}

%HI
Figure~\ref{f:hi} shows a SPIRE color-composite, together with the \hi \ contours from \cite{Zschaechner11b}. The dust emission falls entirely within the second highest \hi \ contour, concentrated in a flat, single disk with no emission in the extended warped \hi \ disk and envelope, in contrast to the more massive NGC 891, where \cite{Popescu03} and \cite{Bianchi11} find evidence for dust throughout the \hi \ disk and the extended envelope \citep{Oosterloo07}. Dust is concentrated in the thin \hi \ disk, but we note that dust and \hi \ clumps do not coincide.

\begin{figure*}
\sidecaption
\includegraphics[width=0.65\textwidth]{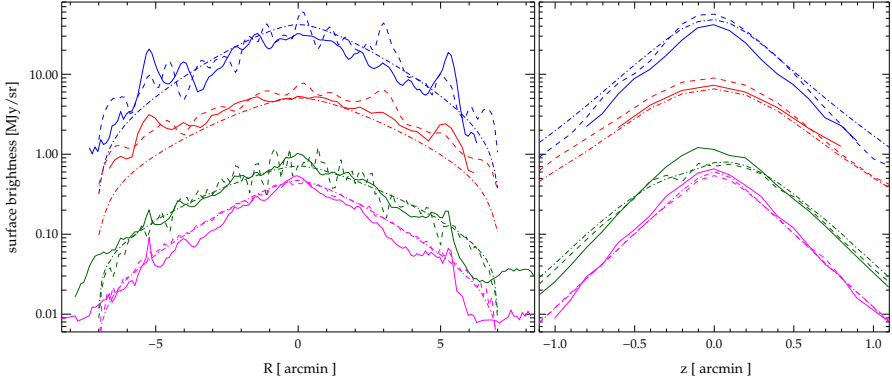}
\caption{\label{f:profiles} Major (left) and minor (right) axis profiles, averaged over a strip of 30" (solid lines), in the {\sc sdss-r} (green) and IRAC 3.6 $\mu$m (pink, offset by a factor 2 for clarity) and the 250 (blue) and 500 (red) $\mu$m SPIRE data, all at native resolution. The major axis profile spans $D_{25}=16\farcm6$ (RC3). We also show the smooth (dot-dash) and clumpy (dashed) TRADING models for comparison (convolved with the SPIRE PSFs at 250 and 500 $\mu$m). We assumed P.A.=$47.3^\circ$ \citep{de-Jong07}.}
\end{figure*}

% profles
Figure \ref{f:profiles} shows the radial and vertical profiles of the sub-mm data, {\sc sdss-r}
data \citep{SDSS-DR8}, and the Spitzer 3.6 \mum\ image from a related Spitzer program\footnote{P.I. R.S. de Jong, Holwerda et al. {\em in preparation}.}.
The sub-mm flux is contained within the $D_{25}$ radius, and the profile shows clumps of
emission at both ends of the disk. % The dusty disk breaks around the same radius (420", 9 kpc) as the break 
% found by \cite{de-Jong07} and the original truncation radius of the stellar disk found by  \cite{vdKruit81a}. 
These clumps of sub-mm emission are the last reliable detection of the disk before the truncation radius \citep[420", 9 kpc][]{vdKruit81a}.
% 
% The break is preceded in both sides of the disk by a clear clump of emission in all three bands as well as the 
% MIPS 24 $\mu$m image (not shown). The symmetry and amplitude suggest two star-formation regions, possibly part of a ring or a spiral structure.
%
% VERTICAL
The vertical sub-mm profiles are similar in width to those in {\em Spitzer} 3.6 $\mu$m emission, a good tracer of the stellar mass \citep[e.g.][]{Meidt12a}. 
% \cite{Comeron11a} find evidence for a thick and thin stellar disk and the dust appears to be mostly in the thin disk. 
% In contrast, there is a second, thicker vertical component in NGC 891 \citep[][]{Kamphuis07}. 
% \cite{Comeron11a} find evidence for a thin and a thick stellar disk in NGC 4244. 
Figure \ref{f:profiles}  suggest that dust is confined to the thin disk, in contrast to NGC 891, where a second vertical component is observed \citep[][]{Kamphuis07,Bianchi11}, as well as a thin dust lane \citep{Xilouris99, Schechtman-Rook12a,Schechtman-Rook12b}.

% \footnote{In NGC 891, optical studies only find the thin dustlane \citep{Xilouris99, Schechtman-Rook11} and occasional chimneys \citep{Howk97}.}. 

% HI-submm

\begin{figure}
\begin{center}
\includegraphics[width=0.49\textwidth]{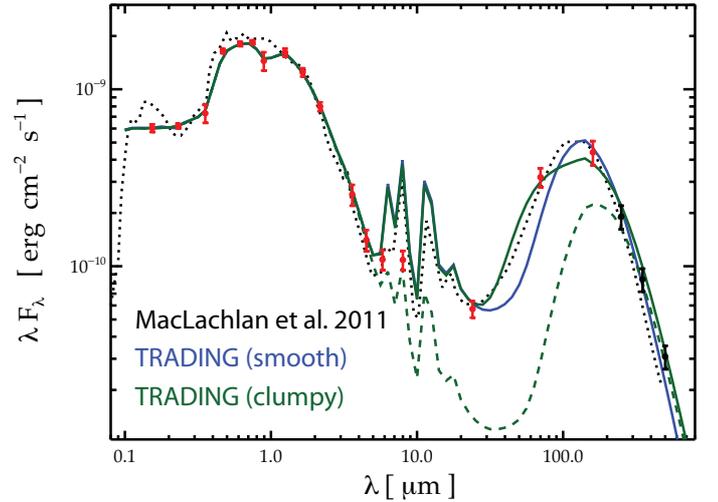}

\caption{\label{f:sedfinal} The Spectral Energy Distribution of NGC 4244. The {\em Herschel/SPIRE} fluxes from this work (black points) are show together with literature data \citep[red symbols,][]{Dale09, MacLachlan11}.  We show the smooth model (blue line); the clumpy model with and without embedded stellar emission (solid and dashed green line, respectively), and the model from MacLachlan et al. (black dotted line). }
\end{center}
\end{figure}

% SED
Fig.~\ref{f:sedfinal} shows the optical and infrared SED of the galaxy. The SPIRE flux densities show a spectral index $F_\nu\propto \nu^{1.6}$,
flatter than the more massive NGC 891 ($F_\nu\propto \nu^{2.7}$), and thus indicative of a colder dust temperature (or of different dust properties). When compared to the radiative transfer model of \cite{MacLachlan11}, the observed fluxes at 350 and 500 $\mu$m are higher by a factor two \citep[Figure \ref{f:sedfinal},][]{Holwerda11iau}. 

% Naive model
We model the surface brightness distribution and SED of NGC 4244 using the radiative transfer code TRADING \citep{Bianchi08}.
We initialize the stellar distribution with the scale-length and height ($h_*, z_*$) as derived by \cite{Fry99} (scaled to 4.4 Mpc), 
and adopted an intrinsic SED that, after being processed by dust, reproduces the observed stellar fluxes in the optical and NIR. 
We truncate the model, for both stars and dust, at 420" (9 kpc) since it is to this radius we have reliable fluxes for both.
% The disk inclination is taken from \citep{Zschaechner11b}.

% SMOOTH MODEL
First, we try a {\em smooth} exponential dust disk with the same scale-height ($z_d$) as for stars, and modify the scale-length ($h_d$) and the central V-band face-on optical depth though the whole disk ($\tau_V^{f.o.}$) to approximate both the sub-mm surface brightness profiles (Fig.~\ref{f:profiles}) and the SED (Fig.~\ref{f:sedfinal}). As \cite{MacLachlan11}, we require a dust scale-length larger than the stellar one to match the observed major axis profiles. However, contrary to \cite{MacLachlan11}, we need a higher opacity (and hence dust mass) to match the peak of the SED\footnote{Although the model of \cite{MacLachlan11} includes a distribution of compact sources, with a dust mass about ten times than that of their diffuse disk, its contribution to the peak SED, as estimated from their Fig.~12, is smaller than 15\%.}. 
Ours and \cite{MacLachlan11} models are compared in Table \ref{t:SED}.
While the smooth model can broadly match the peak of the SED and the profiles, it lacks hot dust to explain emission at $70\mu$m. And it under-predicts the 500 $\mu$m flux, implying a lower dust temperature\footnote{Our assumption is that the dust grain population of NGC 4244 and Milky Way are identical but differences in sub-mm flux and optical extinction can be due to different grain distributions as well.}.

% CLUMPY MODEL
Secondly, we consider an exclusively {\em clumpy} disk model; an exponential distribution of Giant Molecular Clouds (GMC), similar to the clumpy part of the smooth+clumpy model in \cite{Bianchi08} with $M_{GMC} = 10^{6.5} M_\odot$, gas-to-dust mass ratio = 100, $R_{GMC} \simeq 100 ~ pc$. \cite{Bianchi08} find the SED does not depend critically on the GMC size distribution. 
This model matches the sub-mm SED well but fails in the FIR (Fig. \ref{f:sedfinal}, dashed line). 
To match the 70 $\mu$m emission, we heat 60\% of the clouds internally with a hot young starcluster (T = 35000 K, $1.56 \times 10^6 L_\odot$, Fig. \ref{f:sedfinal} solid green line), similar to the second dense dust disk with young stars introduced by \cite{Popescu00} to reconcile FIR and sub-mm fluxes. We note that PAH fluxes (e.g., Spitzer 8 \mum) are not yet reproduced by this model but this may be solved with a range of cloud sizes and heating source types in a future iteration.
The resulting heated clumpy model requires a large dust scale-length, $h_d \sim r_{trunc}$, an increase in total dust mass (three times the {\em smooth} model mass) and some 60\% of the GMCs to be internally heated (equivalent to $\sim$12\% $L_{bol}$ of the stellar disk). 

The smooth and partially heated clumpy SED models illustrate how, in the case of a small disk as NGC 4244, high-resolution sub-mm data reveals a dust disk with more cold dust than a FIR-only model, which is distributed more in clumps and throughout the stellar disk (similar scale-length and height as the stellar disk).
A complete fit of the SED (including future {\em Herschel/PACS} observations) will be presented in the context of our complete low-mass edge-on sample.

%
%These initial SED models of NGC 4244 illustrate that the addition of spatially resolved, sub-mm observations 
%
%The above comparison between SED models illustrates how in NGC 4244, the additional, spatially resolved, sub-mm observations reveal dust scales, both vertical and radial, similar or larger than the stellar disk, a large total dust mass, and a very clumpy dust structure.
%% These SED models illustrate that how in the case of NGC 4244, when (spatially resolved) sub-mm observations are added, the full SED implies large dust scales (both vertical and radial), a large total dust mass, and a very clumpy dust structure to match both sub-mm morphology and fluxes.
%% Some of the clumpy dust structure need to be heated by embedded stellar sources to explain both the sub-mm and FIR fluxes.
%We expect to fully fit the SED and multi-wavelength morphology when our data-set on NC 4244 is complete (including {\em Herschel/PACS} high resolution FIR and JCMT sub-mm observations).
%

%
%\begin{figure}
%\begin{center}
%\includegraphics[width=0.49\textwidth, angle=180]{Figures/n4244-250-smooth.eps}
%\includegraphics[width=0.49\textwidth, angle=180]{Figures/n4244-250-maxclump.eps}
%\caption{ \label{f:modim} The TRADING model images at 250 $\mu$m of NGC 4244, the smoothed and truncated model (top) and the clumped model (bottom). The SPIRE observations (Fig. \ref{f:hi}) resemble the clumped model most.  }
%\end{center}
%\end{figure}
%

%\begin{figure}
%\begin{center}
%\includegraphics[width=0.4\textwidth, angle=180]{Figures/n4244-250-data.eps}
%\includegraphics[width=0.4\textwidth, angle=180]{Figures/n4244-250-smooth.eps}
%\includegraphics[width=0.4\textwidth, angle=180]{Figures/n4244-250-maxclump.eps}
%% \includegraphics[width=0.4\textwidth]{Figures/n4244-250-clumpdist.eps}
%\caption{ \label{f:modim} The 250 $\mu$m image of NGC 4244 with our best smooth model, the maximally clumped model and our best model with a range clumps. The two large clouds at the end of the disk are either spiral structure or star-forming regions.}
%% \caption{ \label{f:modim} The 250 $\mu$m image of NGC 4244 with our best smooth and clumped models, best model with a range clumps. The two large clouds at the end of the disk are either spiral structure or star-forming regions.}
%\end{center}
%\end{figure}

\begin{table}
\caption{Results from the SED models.}
\begin{center}
\begin{tabular}{l l l l l l l}
Parameter 			& MacLachlan			& TRADING			& 					& Unit \\
					& et al. 2011			& Smooth  			& Clumped			& \\
\hline
\hline
$h_*$          			& 1.9 				& 2.3					& 2.3					& kpc\\
$z_*$,$z_d$     			& 0.2 				& 0.3					& 0.3					& kpc\\
$h_d$          			& 3.4 				& 4.5 ($2h_*$)			& 9 $=r_{trunc}$		 	& kpc\\
$r_{trunc}$			& --					& 9 					& 9					& kpc\\

$L_*$          			& $ 3.4 \times 10^9$		& $3.9 \times 10^9$ 		& $3.0 \times 10^9$		& $L_\odot $\\
$M_d$          			& $ 2.4 \times 10^6$		& $0.5 \times 10^7$		& $1.4 \times 10^7$		& $M_\odot $\\
$i$           				& $ 84.5^\circ $			& $ 88^\circ $			&$88^\circ$ 			& \\
$\rm L_{cloud}$      		& $ 1.0 \times 10^8$		& --					& $3.7 \times 10^8$ 		& $L_\odot $\\
$\tau_V^{f.0.} $      		&  0.2 				& 0.4					& 1.1					& \\
%$\tau_{eq}$  			& 1.57 				&					& 					& \\
$h_*/z_*$				& 9.5					& 7.5					& 7.5					& 8.0 (3.6 $\mu$m) \\
$h_d/z_d$  			& 17					& 15					& 30					& 10 (8.0 $\mu$m) \\
\hline
\end{tabular}
\end{center}
\label{t:SED}
\end{table}%

\section{Discussion}
\label{s:disc}

% dust mass (H-ATLAS)
Stacked observations in the {\em Herschel} H-ATLAS survey \citep{hatlas} point to specific characteristics ($T_{d} \sim 15 ~ K$, $M_{d} \sim 10^7 M_\odot$) for blue, low-mass spirals \citep[$\sim10^9 M_\odot$ stellar mass][]{Bourne12a}.
% NGC 4244 has a stellar mass of $6 \times 10^9 M_\odot$ \citep{Strickland04} and an \hi \ mass of $2.5 \times 10^9 M_\odot$ \citep{Zschaechner11b}.
NGC 4244's stellar and \hi\ mass are $6$ and $2.5 \times 10^9 M_\odot$ \citep{Strickland04, Zschaechner11b}.
Compared to the H-ATLAS results, our SED dust mass is reasonable and the gas-to-dust ratio (535 -- 170 for the smooth and clumped model respectively) is in line with those found for smaller spiral disks \citep{Leroy09}.

% gas-to-dust ratio
% CO : 1.4\times10^7 ~ M_\odot
% $2.5 \times 10^9 M_\odot$
% $M_{d} = 4.7 -- 1.47 \times 10^7 M_\odot$
% The gas-to-dust ratio we find (G2D = 535 -- 170 for the smooth and clumped model respectively) is in line with those found for smaller spiral disks \citep{Leroy09}.

% the central optical depth
The central optical depth of $\tau_V^{f.o.} \sim 0.45-1.12$ is similar to the face-on optical depths  in occulting galaxies \citep{kw00a,Holwerda07c} or distant galaxy counts \citep{Holwerda05b, Holwerda07a, Holwerda07b}. 
We therefore feel confident that our SED model accurately reflects the true dust mass of NGC 4244. 

The scale-length of the dust in the SED models is consistently much larger to the stellar one; a factor 1.8 to 4.9, depending on the model. 
For NGC 891, a ratio of 1.5-2 is typically found \citep{Xilouris99, Popescu00, Popescu11, Driver07, Bianchi08, Bianchi11}. 
Similarly, \cite{Holwerda09} and Holwerda et al. {\em submitted} report a clumpy, extended and flat dust profile in a similar mass face-on galaxy occulting a distant bright bulge (2MASX J00482185-2507365, near NGC 253). 
These flat radial profiles of dust suggest radial mixing of the ISM to a similar level found in the stellar disks \citep[][Radburn-Smith et al. {\em submitted}]{de-Jong07, Roskar08}.
%a similar-size galaxy, backlit by a background spiral bulge. The dusty disk is clumpy and extended. 
%In our opinion, these two cases suggest that dust is much more radially mixed in smaller galaxies than in massive ones, similar to their stellar population \citep[][{\em submitted}]{de-Jong07, Radburn-Smith12}
%  (see also Radburn-Smith et al {\em in prep}).

% Dalcanton discussion
We find a dust scale-height similar to the stellar one, as do \cite{MacLachlan11}. They attribute this to either the flattening of dust disks \citep{Dalcanton04} or a smaller stellar scale-height due to inefficient dynamical heating of the stellar disk in lower mass galaxies \citep{Bizyaev04}. 
%
% STELLAR DISK
%
To distinguish between these two scenarios, we compare the oblateness ($h/z$) of the stellar and dust disk (Table \ref{t:SED}) to those in Figure 2 in \cite{Dalcanton04}. We also include the oblateness of the 3.6 and 8.0 $\mu$m {\em Spitzer} observations (Holwerda et al {\em in prep.}). We note that the value for $z_*$ of 300 pc is the scale-height of the thin disk found by \cite{Comeron11a} and RGB stars \citep{Seth05a}.
NGC 4244's stellar disk oblateness is at the highest, i.e., flattest value for low-mass disks, similar to the oblateness of massive disks, arguing against the less efficient disk heating compared to massive disks. 
% We note that lower-mass ($\rm v_{rot} < 120 ~ km/s$) disks are equally likely to display a lower oblateness value and NGC 4244's value is at the highest, i.e., flattest-- end, arguing against the inefficient dynamical heating scenario; 
% The value of the stellar disks's oblateness argues against a lower efficiency of dynamical heating as the maximum stellar disk oblateness is the same for lower and higher masses \citep[which is used as an argument for sub--maximal disks][]{Kregel05b}.
%
% DUST DISK
%
The dusty disk oblateness is extremely high, even compared to the PAH emission ($h/z = 10$ at 8 $\mu$m), but this is predominantly because of the large scale-length of the dust disk.
%  an increase in the scale-length of the disk, which argues for some kind of radial mixing scenario.  
%
% The fact that the scale-height is similar to the stellar one, 
The similar stellar and dust scale-height and the best match by clumpy SED model are both consistent with the scenario from \cite{Dalcanton04}; in a low-mass disk such as this one, the ISM does not collapse into a thin plane but remains fractured and spread throughout the height of the stellar disk.

\section{Summary \& Conclusions}
\label{s:concl}

Based on new sub-mm observations with the {\em Herschel Space Observatory} of NGC 4244, and some illustrative models with the TRADING code, we find that:
\begin{itemize}
\item[i.] The cold dust is confined to the inner, flat \hi \ disk (Fig. \ref{f:hi}).
\item[ii.] The dusty ISM is confined to the stellar radius (Fig. \ref{f:profiles}).
\item[iii.] A smooth model under-predicts 70 \& 500 $\mu$m fluxes (Fig. \ref{f:sedfinal}).
\item[iv.] A clumpy dust disk model with a fraction of the clumps heated by embedded sources fits both the general SED (Fig. \ref{f:sedfinal}) and the morphology of the disk (Fig. \ref{f:profiles}) in sub-mm well.
\item[vi.] The inferred scales of stars and dust are consistent with a scenario that includes some radial migration of dust and a vertical ISM scale supported by turbulence (Table \ref{t:SED}).
\end{itemize}

Future improvements of the NGC 4244 SED model includes new FIR ({\em Herschel/PACS}) information to constrain the cold dust structure and the number of GMCs and the stellar type of their heating sources. We plan to put these results in the context of the full NHEMESES and HEROES {\em Herschel} observations to infer if there is indeed a phase-change with mass in the dust structure and additionally to benchmark current SED models.
%Future work on the NHEMESES project will include FIR ({\em Herschel/PACS}) and 450 \& 850 $\mu$m observations ({JCMT/SCUBA-2}). The wealth of multi-wavelength data will enable us to model the disk of NGC 4244 to the degree that the SED of individual regions such as the two peaks at the end of the disk are self-consistently explained. The SED data-cubes from the NHEMESES and HEROES projects can then be used to test the various SED models for edge-on spiral disks. The combined observations will reveal if there is indeed a phase-change in the ISM of spiral disks as the results presented here and those on NGC 891 from \cite{Bianchi11} seem to indicate.
%\section*{Acknowledgements}

The authors thank ESA for the Fellowship Program and operation of the {\em Herschel Space Observatory}, John MacLachlan for his SED data and Laura Zschaechner for her \hi\ map and the anonymous referee for his or her diligent work.

%\bibliographystyle{aa}
%\bibliography{/Users/bholwerd/Desktop/Science/Bib/Bibliography}

\begin{thebibliography}{61}
\expandafter\ifx\csname natexlab\endcsname\relax\def\natexlab#1{#1}\fi

\bibitem[{{Aihara} {et~al.}(2011){Aihara}, {Allende Prieto}, {An}, {Anderson},
  {Aubourg}, {Balbinot}, {Beers}, {Berlind}, {Bickerton}, {Bizyaev}, {Blanton},
  {Bochanski}, {Bolton}, {Bovy}, {Brandt}, {Brinkmann}, {Brown}, {Brownstein},
  {Busca}, {Campbell}, {Carr}, {Chen}, {Chiappini}, {Comparat}, {Connolly},
  {Cortes}, {Croft}, {Cuesta}, {da Costa}, {Davenport}, {Dawson}, {Dhital},
  {Ealet}, {Ebelke}, {Edmondson}, {Eisenstein}, {Escoffier}, {Esposito},
  {Evans}, {Fan}, {Femen{\'{\i}}a Castell{\'a}}, {Font-Ribera}, {Frinchaboy},
  {Ge}, {Gillespie}, {Gilmore}, {Gonz{\'a}lez Hern{\'a}ndez}, {Gott}, {Gould},
  {Grebel}, {Gunn}, {Hamilton}, {Harding}, {Harris}, {Hawley}, {Hearty}, {Ho},
  {Hogg}, {Holtzman}, {Honscheid}, {Inada}, {Ivans}, {Jiang}, {Johnson},
  {Jordan}, {Jordan}, {Kazin}, {Kirkby}, {Klaene}, {Knapp}, {Kneib},
  {Kochanek}, {Koesterke}, {Kollmeier}, {Kron}, {Lampeitl}, {Lang}, {Le Goff},
  {Lee}, {Lin}, {Long}, {Loomis}, {Lucatello}, {Lundgren}, {Lupton}, {Ma},
  {MacDonald}, {Mahadevan}, {Maia}, {Makler}, {Malanushenko}, {Malanushenko},
  {Mandelbaum}, {Maraston}, {Margala}, {Masters}, {McBride}, {McGehee},
  {McGreer}, {M{\'e}nard}, {Miralda-Escud{\'e}}, {Morrison}, {Mullally},
  {Muna}, {Munn}, {Murayama}, {Myers}, {Naugle}, {Fausti Neto}, {Cuong Nguyen},
  {Nichol}, {O'Connell}, {Ogando}, {Olmstead}, {Oravetz}, {Padmanabhan},
  {Palanque-Delabrouille}, {Pan}, {Pandey}, {P{\^a}ris}, {Percival},
  {Petitjean}, {Pfaffenberger}, {Pforr}, {Phleps}, {Pichon}, {Pieri}, {Prada},
  {Price-Whelan}, {Raddick}, {Ramos}, {Reyl{\'e}}, {Rich}, {Richards}, {Rix},
  {Robin}, {Rocha-Pinto}, {Rockosi}, {Roe}, {Rollinde}, {Ross}, {Ross},
  {Rossetto}, {S{\'a}nchez}, {Sayres}, {Schlegel}, {Schlesinger}, {Schmidt},
  {Schneider}, {Sheldon}, {Shu}, {Simmerer}, {Simmons}, {Sivarani}, {Snedden},
  {Sobeck}, {Steinmetz}, {Strauss}, {Szalay}, {Tanaka}, {Thakar}, {Thomas},
  {Tinker}, {Tofflemire}, {Tojeiro}, {Tremonti}, {Vandenberg}, {Vargas
  Maga{\~n}a}, {Verde}, {Vogt}, {Wake}, {Wang}, {Weaver}, {Weinberg}, {White},
  {White}, {Yanny}, {Yasuda}, {Yeche}, \& {Zehavi}}]{SDSS-DR8}
{Aihara}, H., {Allende Prieto}, C., {An}, D., {et~al.} 2011, \apjs, 193, 29

\bibitem[{{Allen} {et~al.}(1986){Allen}, {Atherton}, \& {Tilanus}}]{Allen86}
{Allen}, R.~J., {Atherton}, P.~D., \& {Tilanus}, R.~P.~J. 1986, \nat, 319, 296

\bibitem[{{Alton} {et~al.}(2004){Alton}, {Xilouris}, {Misiriotis}, {Dasyra}, \&
  {Dumke}}]{Alton04}
{Alton}, P.~B., {Xilouris}, E.~M., {Misiriotis}, A., {Dasyra}, K.~M., \&
  {Dumke}, M. 2004, \aap, 425, 109

\bibitem[{{Baes} {et~al.}(2003){Baes}, {Davies}, {Dejonghe}, {Sabatini},
  {Roberts}, {Evans}, {Linder}, {Smith}, \& {de Blok}}]{Baes03}
{Baes}, M., {Davies}, J.~I., {Dejonghe}, H., {et~al.} 2003, \mnras, 343, 1081

\bibitem[{{Baes} {et~al.}(2010){Baes}, {Fritz}, {Gadotti}, {Smith}, {Dunne},
  {da Cunha}, {Amblard}, {Auld}, {Bendo}, {Bonfield}, {Burgarella},
  {Buttiglione}, {Cava}, {Clements}, {Cooray}, {Dariush}, {de Zotti}, {Dye},
  {Eales}, {Frayer}, {Gonzalez-Nuevo}, {Herranz}, {Ibar}, {Ivison}, {Lagache},
  {Leeuw}, {Lopez-Caniego}, {Jarvis}, {Maddox}, {Negrello}, {Micha{\l}owski},
  {Pascale}, {Pohlen}, {Rigby}, {Rodighiero}, {Samui}, {Serjeant}, {Temi},
  {Thompson}, {van der Werf}, {Verma}, \& {Vlahakis}}]{Baes10a}
{Baes}, M., {Fritz}, J., {Gadotti}, D.~A., {et~al.} 2010, \aap, 518, L39+

\bibitem[{{Baes} {et~al.}(2011){Baes}, {Verstappen}, {De Looze}, {Fritz},
  {Saftly}, {Vidal P{\'e}rez}, {Stalevski}, \& {Valcke}}]{Baes11}
{Baes}, M., {Verstappen}, J., {De Looze}, I., {et~al.} 2011, \apjs, 196, 22

\bibitem[{{Bianchi}(2007)}]{Bianchi07}
{Bianchi}, S. 2007, \aap, 471, 765

\bibitem[{{Bianchi}(2008)}]{Bianchi08}
{Bianchi}, S. 2008, \aap, 490, 461

\bibitem[{{Bianchi} \& {Xilouris}(2011)}]{Bianchi11}
{Bianchi}, S. \& {Xilouris}, E.~M. 2011, \aap, 531, L11+

\bibitem[{{Bizyaev} \& {Kajsin}(2004)}]{Bizyaev04}
{Bizyaev}, D. \& {Kajsin}, S. 2004, \apj, 613, 886

\bibitem[{{Boselli} {et~al.}(2010){Boselli}, {Eales}, {Cortese}, {Bendo},
  {Chanial}, {Buat}, {Davies}, {Auld}, {Rigby}, {Baes}, {Barlow}, {Bock},
  {Bradford}, {Castro-Rodriguez}, {Charlot}, {Clements}, {Cormier}, {Dwek},
  {Elbaz}, {Galametz}, {Galliano}, {Gear}, {Glenn}, {Gomez}, {Griffin}, {Hony},
  {Isaak}, {Levenson}, {Lu}, {Madden}, {O'Halloran}, {Okamura}, {Oliver},
  {Page}, {Panuzzo}, {Papageorgiou}, {Parkin}, {Perez-Fournon}, {Pohlen},
  {Rangwala}, {Roussel}, {Rykala}, {Sacchi}, {Sauvage}, {Schulz}, {Schirm},
  {Smith}, {Spinoglio}, {Stevens}, {Symeonidis}, {Vaccari}, {Vigroux},
  {Wilson}, {Wozniak}, {Wright}, \& {Zeilinger}}]{Boselli10}
{Boselli}, A., {Eales}, S., {Cortese}, L., {et~al.} 2010, \pasp, 122, 261

\bibitem[{{Bourne} {et~al.}(2012){Bourne}, {Maddox}, {Dunne}, {Auld}, {Baes},
  {Baldry}, {Bonfield}, {Cooray}, {Croom}, {Dariush}, {de Zotti}, {Driver},
  {Dye}, {Eales}, {Gomez}, {Gonzalez-Nuevo}, {Hopkins}, {Ibar}, {Jarvis},
  {Lapi}, {Madore}, {Michalowski}, {Pohlen}, {Popescu}, {Rigby}, {Seibert},
  {Smith}, {Tuffs}, {van der Werf}, {Brough}, {Buttiglione}, {Cava},
  {Clements}, {Conselice}, {Fritz}, {Hopwood}, {Ivison}, {Jones}, {Kelvin},
  {Liske}, {Loveday}, {Norberg}, {Robotham}, {Rodighiero}, \&
  {Temi}}]{Bourne12a}
{Bourne}, N., {Maddox}, S.~J., {Dunne}, L., {et~al.} 2012, ArXiv
  e-prints/1201.1916

\bibitem[{{Comer{\'o}n} {et~al.}(2011){Comer{\'o}n}, {Knapen}, {Sheth},
  {Regan}, {Hinz}, {Gil de Paz}, {Men{\'e}ndez-Delmestre}, {Mu{\~n}oz-Mateos},
  {Seibert}, {Kim}, {Athanassoula}, {Bosma}, {Buta}, {Elmegreen}, {Ho},
  {Holwerda}, {Laurikainen}, {Salo}, \& {Schinnerer}}]{Comeron11a}
{Comer{\'o}n}, S., {Knapen}, J.~H., {Sheth}, K., {et~al.} 2011, \apj, 729, 18

\bibitem[{{Dalcanton} {et~al.}(2004){Dalcanton}, {Yoachim}, \&
  {Bernstein}}]{Dalcanton04}
{Dalcanton}, J.~J., {Yoachim}, P., \& {Bernstein}, R.~A. 2004, \apj, 608, 189

\bibitem[{{Dale} {et~al.}(2009){Dale}, {Cohen}, {Johnson}, {Schuster},
  {Calzetti}, {Engelbracht}, {Gil de Paz}, {Kennicutt}, {Lee}, {Begum},
  {Block}, {Dalcanton}, {Funes}, {Gordon}, {Johnson}, {Marble}, {Sakai},
  {Skillman}, {van Zee}, {Walter}, {Weisz}, {Williams}, {Wu}, \& {Wu}}]{Dale09}
{Dale}, D.~A., {Cohen}, S.~A., {Johnson}, L.~C., {et~al.} 2009, \apj, 703, 517

\bibitem[{{Davies} {et~al.}(2010){Davies}, {Baes}, {Bendo}, {Bianchi},
  {Bomans}, {Boselli}, {Clemens}, {Corbelli}, {Cortese}, {Dariush}, {de Looze},
  {di Serego Alighieri}, {Fadda}, {Fritz}, {Garcia-Appadoo}, {Gavazzi},
  {Giovanardi}, {Grossi}, {Hughes}, {Hunt}, {Jones}, {Madden}, {Pierini},
  {Pohlen}, {Sabatini}, {Smith}, {Verstappen}, {Vlahakis}, {Xilouris}, \&
  {Zibetti}}]{Davies10a}
{Davies}, J.~I., {Baes}, M., {Bendo}, G.~J., {et~al.} 2010, \aap, 518, L48+

\bibitem[{{de Jong} {et~al.}(2007){de Jong}, {Seth}, {Radburn-Smith}, {Bell},
  {Brown}, {Bullock}, {Courteau}, {Dalcanton}, {Ferguson}, {Goudfrooij},
  {Holfeltz}, {Holwerda}, {Purcell}, {Sick}, \& {Zucker}}]{de-Jong07}
{de Jong}, R.~S., {Seth}, A.~C., {Radburn-Smith}, D.~J., {et~al.} 2007, \apjl,
  667, L49

\bibitem[{{Driver} {et~al.}(2007){Driver}, {Popescu}, {Tuffs}, {Liske},
  {Graham}, {Allen}, \& {de Propris}}]{Driver07}
{Driver}, S.~P., {Popescu}, C.~C., {Tuffs}, R.~J., {et~al.} 2007, \mnras, 563

\bibitem[{{Eales} {et~al.}(2010{\natexlab{a}}){Eales}, {Dunne}, {Clements},
  {Cooray}, {de Zotti}, {Dye}, {Ivison}, {Jarvis}, {Lagache}, {Maddox},
  {Negrello}, {Serjeant}, {Thompson}, {van Kampen}, {Amblard}, {Andreani},
  {Baes}, {Beelen}, {Bendo}, {Benford}, {Bertoldi}, {Bock}, {Bonfield},
  {Boselli}, {Bridge}, {Buat}, {Burgarella}, {Carlberg}, {Cava}, {Chanial},
  {Charlot}, {Christopher}, {Coles}, {Cortese}, {Dariush}, {da Cunha},
  {Dalton}, {Danese}, {Dannerbauer}, {Driver}, {Dunlop}, {Fan}, {Farrah},
  {Frayer}, {Frenk}, {Geach}, {Gardner}, {Gomez}, {Gonz{\'a}lez-Nuevo},
  {Gonz{\'a}lez-Solares}, {Griffin}, {Hardcastle}, {Hatziminaoglou}, {Herranz},
  {Hughes}, {Ibar}, {Jeong}, {Lacey}, {Lapi}, {Lawrence}, {Lee}, {Leeuw},
  {Liske}, {L{\'o}pez-Caniego}, {M{\"u}ller}, {Nandra}, {Panuzzo},
  {Papageorgiou}, {Patanchon}, {Peacock}, {Pearson}, {Phillipps}, {Pohlen},
  {Popescu}, {Rawlings}, {Rigby}, {Rigopoulou}, {Robotham}, {Rodighiero},
  {Sansom}, {Schulz}, {Scott}, {Smith}, {Sibthorpe}, {Smail}, {Stevens},
  {Sutherland}, {Takeuchi}, {Tedds}, {Temi}, {Tuffs}, {Trichas}, {Vaccari},
  {Valtchanov}, {van der Werf}, {Verma}, {Vieria}, {Vlahakis}, \&
  {White}}]{hatlas}
{Eales}, S., {Dunne}, L., {Clements}, D., {et~al.} 2010{\natexlab{a}}, \pasp,
  122, 499

\bibitem[{{Eales} {et~al.}(2010{\natexlab{b}}){Eales}, {Smith}, {Wilson},
  {Bendo}, {Cortese}, {Pohlen}, {Boselli}, {Gomez}, {Auld}, {Baes}, {Barlow},
  {Bock}, {Bradford}, {Buat}, {Castro-Rodr{\'{\i}}guez}, {Chanial}, {Charlot},
  {Ciesla}, {Clements}, {Cooray}, {Cormier}, {Davies}, {Dwek}, {Elbaz},
  {Galametz}, {Galliano}, {Gear}, {Glenn}, {Griffin}, {Hony}, {Isaak},
  {Levenson}, {Lu}, {Madden}, {O'Halloran}, {Okumura}, {Oliver}, {Page},
  {Panuzzo}, {Papageorgiou}, {Parkin}, {P{\'e}rez-Fournon}, {Rangwala},
  {Rigby}, {Roussel}, {Rykala}, {Sacchi}, {Sauvage}, {Schulz}, {Schirm},
  {Spinoglio}, {Srinivasan}, {Stevens}, {Symeonidis}, {Trichas}, {Vaccari},
  {Vigroux}, {Wozniak}, {Wright}, \& {Zeilinger}}]{Eales10}
{Eales}, S.~A., {Smith}, M.~W.~L., {Wilson}, C.~D., {et~al.}
  2010{\natexlab{b}}, \aap, 518, L62+

\bibitem[{{Florido} {et~al.}(1991){Florido}, {Battaner}, {Sanchez-Saavedra},
  {Prieto}, \& {Mediavilla}}]{Florido91a}
{Florido}, E., {Battaner}, E., {Sanchez-Saavedra}, M.~L., {Prieto}, M., \&
  {Mediavilla}, E. 1991, \mnras, 251, 193

\bibitem[{{Fry} {et~al.}(1999){Fry}, {Morrison}, {Harding}, \&
  {Boroson}}]{Fry99}
{Fry}, A.~M., {Morrison}, H.~L., {Harding}, P., \& {Boroson}, T.~A. 1999, \aj,
  118, 1209

\bibitem[{{Gordon} {et~al.}(2001){Gordon}, {Misselt}, {Witt}, \&
  {Clayton}}]{Gordon01}
{Gordon}, K.~D., {Misselt}, K.~A., {Witt}, A.~N., \& {Clayton}, G.~C. 2001,
  \apj, 551, 269

\bibitem[{{Griffin} {et~al.}(2010){Griffin}, {Abergel}, {Abreu}, {Ade},
  {Andr{\'e}}, {Augueres}, {Babbedge}, {Bae}, {Baillie}, {Baluteau}, {Barlow},
  {Bendo}, {Benielli}, {Bock}, {Bonhomme}, {Brisbin}, {Brockley-Blatt},
  {Caldwell}, {Cara}, {Castro-Rodriguez}, {Cerulli}, {Chanial}, {Chen},
  {Clark}, {Clements}, {Clerc}, {Coker}, {Communal}, {Conversi}, {Cox},
  {Crumb}, {Cunningham}, {Daly}, {Davis}, {de Antoni}, {Delderfield}, {Devin},
  {di Giorgio}, {Didschuns}, {Dohlen}, {Donati}, {Dowell}, {Dowell}, {Duband},
  {Dumaye}, {Emery}, {Ferlet}, {Ferrand}, {Fontignie}, {Fox}, {Franceschini},
  {Frerking}, {Fulton}, {Garcia}, {Gastaud}, {Gear}, {Glenn}, {Goizel},
  {Griffin}, {Grundy}, {Guest}, {Guillemet}, {Hargrave}, {Harwit}, {Hastings},
  {Hatziminaoglou}, {Herman}, {Hinde}, {Hristov}, {Huang}, {Imhof}, {Isaak},
  {Israelsson}, {Ivison}, {Jennings}, {Kiernan}, {King}, {Lange}, {Latter},
  {Laurent}, {Laurent}, {Leeks}, {Lellouch}, {Levenson}, {Li}, {Li},
  {Lilienthal}, {Lim}, {Liu}, {Lu}, {Madden}, {Mainetti}, {Marliani}, {McKay},
  {Mercier}, {Molinari}, {Morris}, {Moseley}, {Mulder}, {Mur}, {Naylor},
  {Nguyen}, {O'Halloran}, {Oliver}, {Olofsson}, {Olofsson}, {Orfei}, {Page},
  {Pain}, {Panuzzo}, {Papageorgiou}, {Parks}, {Parr-Burman}, {Pearce},
  {Pearson}, {P{\'e}rez-Fournon}, {Pinsard}, {Pisano}, {Podosek}, {Pohlen},
  {Polehampton}, {Pouliquen}, {Rigopoulou}, {Rizzo}, {Roseboom}, {Roussel},
  {Rowan-Robinson}, {Rownd}, {Saraceno}, {Sauvage}, {Savage}, {Savini},
  {Sawyer}, {Scharmberg}, {Schmitt}, {Schneider}, {Schulz}, {Schwartz},
  {Shafer}, {Shupe}, {Sibthorpe}, {Sidher}, {Smith}, {Smith}, {Smith},
  {Spencer}, {Stobie}, {Sudiwala}, {Sukhatme}, {Surace}, {Stevens}, {Swinyard},
  {Trichas}, {Tourette}, {Triou}, {Tseng}, {Tucker}, {Turner}, {Vaccari},
  {Valtchanov}, {Vigroux}, {Virique}, {Voellmer}, {Walker}, {Ward}, {Waskett},
  {Weilert}, {Wesson}, {White}, {Whitehouse}, {Wilson}, {Winter}, {Woodcraft},
  {Wright}, {Xu}, {Zavagno}, {Zemcov}, {Zhang}, \& {Zonca}}]{Griffin10}
{Griffin}, M.~J., {Abergel}, A., {Abreu}, A., {et~al.} 2010, \aap, 518, L3+

\bibitem[{{Heald} {et~al.}(2011{\natexlab{a}}){Heald}, {Allan}, {Zschaechner},
  {Kamphuis}, {Rand}, {J{\'o}zsa}, \& {Gentile}}]{Heald11b}
{Heald}, G., {Allan}, J., {Zschaechner}, L., {et~al.} 2011{\natexlab{a}}, in
  IAU Symposium, Vol. 277, IAU Symposium, ed. {C.~Carignan, F.~Combes, \&
  K.~C.~Freeman}, 59--62

\bibitem[{{Heald} {et~al.}(2011{\natexlab{b}}){Heald}, {J{\'o}zsa}, {Serra},
  {Zschaechner}, {Rand}, {Fraternali}, {Oosterloo}, {Walterbos}, {J{\"u}tte},
  \& {Gentile}}]{Heald11a}
{Heald}, G., {J{\'o}zsa}, G., {Serra}, P., {et~al.} 2011{\natexlab{b}}, \aap,
  526, A118

\bibitem[{{Holwerda} {et~al.}(2011){Holwerda}, {Bianchi}, {Baes}, {de Jong},
  {Dalcanton}, {Radburn-Smith}, {Gordon}, \& {Xilouris}}]{Holwerda11iau}
{Holwerda}, B.~W., {Bianchi}, S., {Baes}, M., {et~al.} 2011, ArXiv :1109.5603

\bibitem[{{Holwerda} {et~al.}(2007{\natexlab{a}}){Holwerda}, {Draine},
  {Gordon}, {Gonz\'alez}, {Calzetti}, {Thornley}, {Buckalew}, {Allen}, \& {van
  der Kruit}}]{Holwerda07a}
{Holwerda}, B.~W., {Draine}, B., {Gordon}, K.~D., {et~al.} 2007{\natexlab{a}},
  \aj, 134, 2226

\bibitem[{{Holwerda} {et~al.}(2005){Holwerda}, {Gonz\'alez}, {Allen}, \& {van
  der Kruit}}]{Holwerda05b}
{Holwerda}, B.~W., {Gonz\'alez}, R.~A., {Allen}, R.~J., \& {van der Kruit},
  P.~C. 2005, \aj, 129, 1396

\bibitem[{{Holwerda} {et~al.}(2007{\natexlab{b}}){Holwerda}, {Keel}, \&
  {Bolton}}]{Holwerda07c}
{Holwerda}, B.~W., {Keel}, W.~C., \& {Bolton}, A. 2007{\natexlab{b}}, \aj, 134,
  2385

\bibitem[{{Holwerda} {et~al.}(2009){Holwerda}, {Keel}, {Williams}, {Dalcanton},
  \& {de Jong}}]{Holwerda09}
{Holwerda}, B.~W., {Keel}, W.~C., {Williams}, B., {Dalcanton}, J.~J., \& {de
  Jong}, R.~S. 2009, \aj, 137, 3000

\bibitem[{{Holwerda} {et~al.}(2007{\natexlab{c}}){Holwerda}, {Meyer}, {Regan},
  {Calzetti}, {Gordon}, {Smith}, {Dale}, {Engelbracht}, {Jarrett}, {Thornley},
  {Bot}, {Buckalew}, {Kennicutt}, \& {Gonz{\'a}lez}}]{Holwerda07b}
{Holwerda}, B.~W., {Meyer}, M., {Regan}, M., {et~al.} 2007{\natexlab{c}}, \aj,
  134, 1655

\bibitem[{{Hoopes} {et~al.}(1999){Hoopes}, {Walterbos}, \& {Rand}}]{Hoopes99}
{Hoopes}, C.~G., {Walterbos}, R.~A.~M., \& {Rand}, R.~J. 1999, \apj, 522, 669

\bibitem[{{Kamphuis} {et~al.}(2007){Kamphuis}, {Holwerda}, {Allen}, {Peletier},
  \& {van der Kruit}}]{Kamphuis07}
{Kamphuis}, P., {Holwerda}, B.~W., {Allen}, R.~J., {Peletier}, R.~F., \& {van
  der Kruit}, P.~C. 2007, \aap, 471, L1

\bibitem[{{Kodaira} \& {Yamashita}(1996)}]{Kodaira96}
{Kodaira}, K. \& {Yamashita}, T. 1996, \pasj, 48, 581

\bibitem[{{Leroy} {et~al.}(2009){Leroy}, {Walter}, {Bigiel}, {Usero}, {Weiss},
  {Brinks}, {de Blok}, {Kennicutt}, {Schuster}, {Kramer}, {Wiesemeyer}, \&
  {Roussel}}]{Leroy09}
{Leroy}, A.~K., {Walter}, F., {Bigiel}, F., {et~al.} 2009, \aj, 137, 4670

\bibitem[{{MacLachlan} {et~al.}(2011){MacLachlan}, {Matthews}, {Wood}, \&
  {Gallagher}}]{MacLachlan11}
{MacLachlan}, J.~M., {Matthews}, L.~D., {Wood}, K., \& {Gallagher}, J.~S. 2011,
  \apj, 741, 6

\bibitem[{{Matthews} \& {Wood}(2001)}]{Matthews01}
{Matthews}, L.~D. \& {Wood}, K. 2001, \apj, 548, 150

\bibitem[{{Meidt} {et~al.}(2012){Meidt}, {Schinnerer}, {Knapen}, {Bosma},
  {Athanassoula}, {Sheth}, {Buta}, {Zaritsky}, {Laurikainen}, {Elmegreen},
  {Elmegreen}, {Gadotti}, {Salo}, {Regan}, {Ho}, {Madore}, {Hinz}, {Skibba},
  {Gil de Paz}, {Mu{\~n}oz-Mateos}, {Men{\'e}ndez-Delmestre}, {Seibert}, {Kim},
  {Mizusawa}, {Laine}, \& {Comer{\'o}n}}]{Meidt12a}
{Meidt}, S.~E., {Schinnerer}, E., {Knapen}, J.~H., {et~al.} 2012, \apj, 744, 17

\bibitem[{{Misselt} {et~al.}(2001){Misselt}, {Gordon}, {Clayton}, \&
  {Wolff}}]{Misselt01}
{Misselt}, K.~A., {Gordon}, K.~D., {Clayton}, G.~C., \& {Wolff}, M.~J. 2001,
  \apj, 551, 277

\bibitem[{{Olling}(1996)}]{Olling96b}
{Olling}, R.~P. 1996, \aj, 112, 457

\bibitem[{{Oosterloo} {et~al.}(2007){Oosterloo}, {Fraternali}, \&
  {Sancisi}}]{Oosterloo07}
{Oosterloo}, T., {Fraternali}, F., \& {Sancisi}, R. 2007, \aj, 134, 1019

\bibitem[{{Ott}(2010)}]{Ott10}
{Ott}, S. 2010, in Astronomical Society of the Pacific Conference Series, Vol.
  434, Astronomical Data Analysis Software and Systems XIX, ed. {Y.~Mizumoto,
  K.-I.~Morita, \& M.~Ohishi}, 139

\bibitem[{{Pierini} {et~al.}(2004){Pierini}, {Gordon}, {Witt}, \&
  {Madsen}}]{Pierini04}
{Pierini}, D., {Gordon}, K.~D., {Witt}, A.~N., \& {Madsen}, G.~J. 2004, \apj,
  617, 1022

\bibitem[{{Pilbratt} {et~al.}(2010){Pilbratt}, {Riedinger}, {Passvogel},
  {Crone}, {Doyle}, {Gageur}, {Heras}, {Jewell}, {Metcalfe}, {Ott}, \&
  {Schmidt}}]{Pilbratt10a}
{Pilbratt}, G.~L., {Riedinger}, J.~R., {Passvogel}, T., {et~al.} 2010, \aap,
  518, L1+

\bibitem[{{Popescu} {et~al.}(2000){Popescu}, {Misiriotis}, {Kylafis}, {Tuffs},
  \& {Fischera}}]{Popescu00}
{Popescu}, C.~C., {Misiriotis}, A., {Kylafis}, N.~D., {Tuffs}, R.~J., \&
  {Fischera}, J. 2000, \aap, 362, 138

\bibitem[{{Popescu} \& {Tuffs}(2003)}]{Popescu03}
{Popescu}, C.~C. \& {Tuffs}, R.~J. 2003, \aap, 410, L21

\bibitem[{{Popescu} {et~al.}(2011){Popescu}, {Tuffs}, {Dopita}, {Fischera},
  {Kylafis}, \& {Madore}}]{Popescu11}
{Popescu}, C.~C., {Tuffs}, R.~J., {Dopita}, M.~A., {et~al.} 2011, \aap, 527,
  A109+

\bibitem[{{Radburn-Smith} {et~al.}(2011){Radburn-Smith}, {de Jong}, {Seth},
  {Bailin}, {Bell}, {Brown}, {Bullock}, {Courteau}, {Dalcanton}, {Ferguson},
  {Goudfrooij}, {Holfeltz}, {Holwerda}, {Purcell}, {Sick}, {Streich}, {Vlajic},
  \& {Zucker}}]{GHOSTS}
{Radburn-Smith}, D.~J., {de Jong}, R.~S., {Seth}, A.~C., {et~al.} 2011, \apjs,
  195, 18

\bibitem[{{Ro{\v s}kar} {et~al.}(2008){Ro{\v s}kar}, {Debattista}, {Stinson},
  {Quinn}, {Kaufmann}, \& {Wadsley}}]{Roskar08}
{Ro{\v s}kar}, R., {Debattista}, V.~P., {Stinson}, G.~S., {et~al.} 2008, \apjl,
  675, L65

\bibitem[{{Schechtman-Rook} {et~al.}(2012{\natexlab{a}}){Schechtman-Rook},
  {Bershady}, \& {Wood}}]{Schechtman-Rook12a}
{Schechtman-Rook}, A., {Bershady}, M.~A., \& {Wood}, K. 2012{\natexlab{a}},
  \apj, 746, 70

\bibitem[{{Schechtman-Rook} {et~al.}(2012{\natexlab{b}}){Schechtman-Rook},
  {Bershady}, {Wood}, \& {Robitaille}}]{Schechtman-Rook12b}
{Schechtman-Rook}, A., {Bershady}, M.~A., {Wood}, K., \& {Robitaille}, T.~P.
  2012{\natexlab{b}}, ArXiv e-prints/1203.0023

\bibitem[{{Seth} {et~al.}(2008){Seth}, {Ag{\"u}eros}, {Lee}, \&
  {Basu-Zych}}]{Seth08}
{Seth}, A., {Ag{\"u}eros}, M., {Lee}, D., \& {Basu-Zych}, A. 2008, \apj, 678,
  116

\bibitem[{{Seth} {et~al.}(2005){Seth}, {Dalcanton}, \& {de Jong}}]{Seth05a}
{Seth}, A.~C., {Dalcanton}, J.~J., \& {de Jong}, R.~S. 2005, \aj, 129, 1331

\bibitem[{{Strickland} {et~al.}(2004){Strickland}, {Heckman}, {Colbert},
  {Hoopes}, \& {Weaver}}]{Strickland04}
{Strickland}, D.~K., {Heckman}, T.~M., {Colbert}, E.~J.~M., {Hoopes}, C.~G., \&
  {Weaver}, K.~A. 2004, \apjs, 151, 193

\bibitem[{{van der Kruit} \& {Searle}(1981)}]{vdKruit81a}
{van der Kruit}, P.~C. \& {Searle}, L. 1981, \aap, 95, 105

\bibitem[{{Weingartner} \& {Draine}(2001)}]{Weingartner01b}
{Weingartner}, J.~C. \& {Draine}, B.~T. 2001, \apj, 553, 581

\bibitem[{{White} {et~al.}(2000){White}, {Keel}, \& {Conselice}}]{kw00a}
{White}, III, R.~E., {Keel}, W.~C., \& {Conselice}, C.~J. 2000, \apj, 542, 761

\bibitem[{{Xilouris} {et~al.}(1999){Xilouris}, {Byun}, {Kylafis}, {Paleologou},
  \& {Papamastorakis}}]{Xilouris99}
{Xilouris}, E.~M., {Byun}, Y.~I., {Kylafis}, N.~D., {Paleologou}, E.~V., \&
  {Papamastorakis}, J. 1999, \aap, 344, 868

\bibitem[{{Yoachim} \& {Dalcanton}(2006)}]{Yoachim06}
{Yoachim}, P. \& {Dalcanton}, J.~J. 2006, \aj, 131, 226

\bibitem[{{Zschaechner} {et~al.}(2011){Zschaechner}, {Rand}, {Heald},
  {Gentile}, \& {Kamphuis}}]{Zschaechner11b}
{Zschaechner}, L.~K., {Rand}, R.~J., {Heald}, G.~H., {Gentile}, G., \&
  {Kamphuis}, P. 2011, \apj, 740, 35

\end{thebibliography}

\end{document}